\documentclass[runningheads]{llncs}
\usepackage{url}
\usepackage{graphicx}
\usepackage{subfig}

\begin{document}
\title{An Interface between Legacy and Modern Mobile Devices for Digital Identity}
\titlerunning{Interfacing between Legacy and Modern Mobile Devices}
\author{Vasilios Mavroudis\inst{1} \and
Chris Hicks\inst{1}\and
Jon Crowcroft\inst{1,2}}
\authorrunning{V. Mavroudis et al.}

\institute{Alan Turing Institute \and University of Cambridge\\
\email{\{vmavroudis,c.hicks, jcrowcroft\}@turing.ac.uk}}
\maketitle              % typeset the header of the contribution
\begin{abstract}
In developing regions a substantial number of users rely on legacy and ultra-low-cost mobile devices. Unfortunately, many of these devices are not equipped to run the standard authentication or identity apps that are available for smartphones. Increasingly, apps that display Quick Response (QR) codes are being used to communicate personal credentials (e.g., Covid-19 vaccination certificates). This paper describes a novel interface for QR code credentials that is compatible with legacy mobile devices. Our solution, which we have released under open source licensing, allows Web Application Enabled legacy mobile devices to load and display standard QR codes. This technique makes modern identity platforms available to previously excluded and economically disadvantaged populations.

\end{abstract}

\section{Introduction}\label{sec:intro}
Covid-19 has accelerated the push for digital identity solutions globally. The commercial and government pressure for universal Covid-19 vaccine certificates has yielded a range of new technologies for displaying personal details using smartphone devices. The de facto approach has become to display a Quick Response (QR) code containing either cryptographically signed personal attributes or a Uniform Resource Identifier (URI) to an online service providing the same information. Consideration has also been given to groups without access to smartphones, usually in the form of printable QR codes which simply reproduce what would otherwise be shown on a device screen. Amongst all of the existing work however, little consideration has been given to the needs of users with legacy or ultra-low-cost mobile devices. Although these devices, also known as feature phones, are uncommon in many developed countries they are still mainstream in many parts of the world. In Sub-Saharan Africa, for example, an estimated 56\% of all cellular connections in 2019 were made from feature phones~\cite{gsma_africa}. Innovative applications for these legacy devices currently range from mobile banking and secure payments~\cite{Huges07_MPesa,baqer_digitally_17}, now an integral part of the economy in Kenya~\cite{jack2011mobile}, to supporting agriculture and farming~\cite{kenya_farming_Krell21} but have not yet widely support digital identity. 

The main barrier to using feature phones for digital identity is the absence of a standard developer-friendly mobile operating system. Unlike smartphones which usually run either the Google Android or Apple iOS operating systems (OSes), most feature phones run proprietary, closed-source and even discontinued OSes (e.g., the Nokia Series 30~\cite{Nokia_S30WAP_03} or the Sony Ericsson A1/200). This means that unless an identity application is installed by the manufacturer it is nearly impossible to deploy one to new users. Without a programmable OS which is developer-friendly, existing identity solutions are limited to either programming the Subscriber Identity Module (SIM) card~\cite{imza_krimpe_14,mobile_id_Laud_09} or receiving One-Time Passcodes (OTPs) via SMS; in both cases, the user token is restricted to a few digits which are displayed on the screen. 

In this work we demonstrate a robust and novel method of displaying QR codes for identification using legacy mobile devices. In particular, the approach we propose:
\begin{itemize}
    \item Enables previously overlooked legacy mobile devices with hundreds of millions of users to benefit from modern digital identity solutions. 
    
    \item Shows how standard QR code certificates can be retrofitted to low-cost mobile devices with very limited interfacing capabilities.
    An important feature is that our approach is able to tolerate intermittent user connectivity by exploiting local caching.
    
    \item Is freely available with an implementation and documentation which we make public under the MIT open-source license.
\end{itemize}

\section{Background}\label{sec:background}
We now outline the fundamental concepts and technologies used in the rest of this work.

\subsection{Wireless Application Protocol}
\begin{figure}
\caption{Illustration of a webpage load under the WAP protocol. The user's device does not have access to an IP network but
instead submits its requests through the WAP Gateway which fetches and encodes the WML page from the third-party server.}
\centering
\includegraphics[width=1\textwidth]{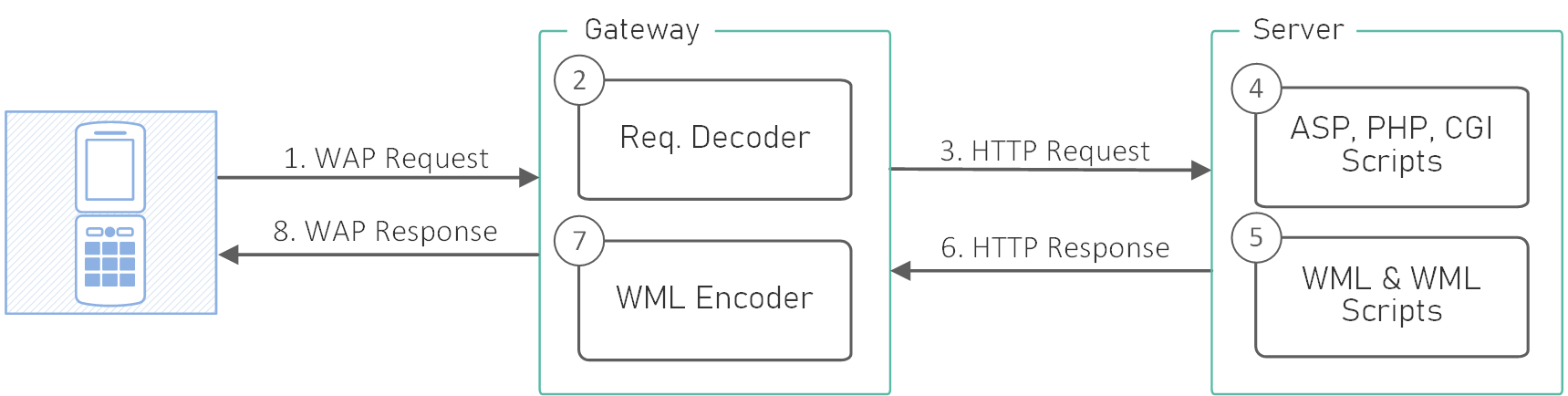}
\label{fig:wap}
\end{figure}

The Wireless Application Protocol (WAP) was introduced in 1999~\cite{sharma2003wireless} and offers a way for mobile devices 
(such as feature phones) to access information over a wireless network. Unlike modern devices, feature phones are not able to 
render HTML pages, execute Javascript or submit HTTP requests. Instead, compatible servers serve content in the Wireless Markup
Language (WML) and WML Script, while the only graphics format supported is Wireless Application Protocol Bitmap Format (WBMP).
WAP uses a gateway (Figure~\ref{fig:wap}) that decodes the WAP requests (sent from the phone's microbrowser) into HTTP 
requests and routes them to the online server. Similarly, the HTTP responses are first encoded into a binary representation of WML 
by the gateway and then forwarded to the phone.

\subsection{QR codes for Digital Identity}
QR codes are a form of machine-readable barcode originally intended to support traceability in automotive supply chains \cite{qr_patent}.  Contemporary standards, such as those intended to provide interoperable vaccination information \cite{EU_Covid_Factsheet,Vaccine_Cert_Interim}, have seen a re-popularisation of QR codes as a mechanism for encoding personal identity information. QR codes are composed of modules and the number of modules dictates both the physical size and information capacity. There are standard module configurations for QR codes, termed the version number, which range from 21x21 modules for a version 1 code to 177x177 modules for a version 40 code~\cite{QR_ISO}. Since a module needs to be at least (and divisible by) 1x1 pixels in size, we can determine the maximum information capacity of a QR code based on a mobile device's screen size. Our testing using a Nokia 3510 device, which has a total screen size of 96x64, indicates that 46x46 pixels are available within the WAP browser to display a QR code. As shown in Figure~\ref{fig:phone}, this reduction arises because several pixels from the total available screen size are required for the WAP browser user interface provided by the mobile device. Allowing for an Error Correcting Code (ECC) level of 7\%, a 46x46 pixel space bounds the maximum QR code information capacity to the 224 alphanumeric characters provided by a version 7 code. In practice, the available information capacity is reduced as a suitable encoding must be used to represent binary data. The de facto standard encoding being used in a number of health certification applications is base 45~\cite{base_45} and this permits storing 2 binary bytes per 3 alphanumeric characters. In other words, a version 7 QR code can encode around 149 bytes of uncompressed information and this is enough to include standard public-key digital signatures using modest curve sizes.

\section{Problem Statement}
As outlined in Section~\ref{sec:intro}, most applications assume devices with modern interfacing capabilities during the execution of their authentication protocols. Our goal is to enable feature phone users to participate in modern authentication protocols, without any additional investment in equipment.

In the rest of this work, we refer to the feature phone user as \textit{Prover}. The Prover wishes to prove their identity 
or an attribute (e.g., their age) to the \textit{Verifier}. The Verifier can be a government official, a local merchant, 
a bank representative or a local organisation running a food subsidy. Besides these actors, we also introduce an external \textit{Service} that is available over the Internet and is tasked with keeping records of the users' details and can issue \textit{Tokens}. Tokens encapsulate various types of data such as digital certificates, signatures, URIs and plain user information depending on the use case.

\begin{figure}
\caption{\textbf{Setup.} The Prover wishes to prove their identity or status to the Verifier. The Verifier uses a modern device, while the Prover possesses a feature phone, making it hard to interface with nearby devices (dotted arrow). Both the Prover and Verifier are \textit{intermittently} connected to the Internet.}
\centering
\includegraphics[width=0.55\textwidth]{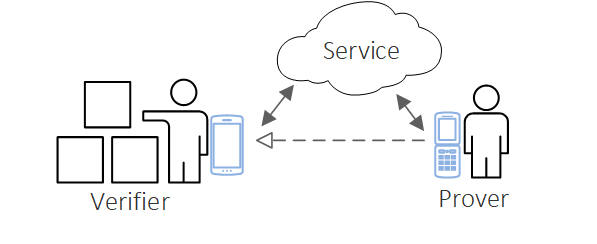}
\label{fig:setup}
\end{figure}

As seen in Figure~\ref{fig:setup}, both the Prover and the Verifier need to be able to connect to the Internet using their devices.
This connectivity may be intermittent for one or both parties.
In line with the sparse mobile network coverage seen in many rural and less developed regions, their connectivity is assumed to intermittent.
Moreover, the Prover is able to access online services only through low-bandwidth WAP due to the limitations of their device. 
The Verifier's device needs to be capable of scanning QR codes. A basic smartphone or a considerably cheaper smart-feature phone
could be used.

\subsection{Use Cases}\label{sec:usecases}
We now outline two use cases that require a cross-device interface.
Both of them require the Prover and the Verifier to be in physical proximity.\\

\noindent\textit{1. Identity or Status Credentials.}
The Prover wants to convince the Verifier of their identity or an attribute/status (e.g., over 18).
From their end, the Verifier needs to be presented with adequate proof that the claims of the Prover are truthful. 
We assume that the two parties do not necessarily have access to the Internet during the execution of the protocol.
The implementation of the Covid-19 certificates in the EU~\cite{Vaccine_Cert_Interim} is a straightforward instantiation of this use case.\\

\noindent\textit{2. Food Subsidy Management.}
The Prover wants to convince the Verifier that they can receive the food subsidy. 
The Verifier needs to be presented with adequate proof that the Prover is eligible for the subsidy and that they have not already used their entitlement. 

\subsection{Threat Model}
A Prover can be malicious and may attempt to claim attributes or request access 
to services they are not entitled to. We assume that provers can collude with other 
malicious users and are byzantine i.e., may deviate from the correct execution of 
the protocol. For example, such an adversary may attempt to craft malicious tokens, replay 
tokens from other users or modify their own past tokens. The Verifier is also potentially 
malicious and may, for instance, capture credentials submitted by users and attempt to replay 
them to other victim verifiers.

We also assume that the Service is trustworthy and that the provider does not act maliciously  
(either independently or by colluding with other malicious parties). All online communications are 
encrypted using standard encryption techniques (e.g., TLS 1.3~\cite{rescorla2018transport}).
Denial of service attacks from the Verifier or the service towards provers are not within the
scope of this work. Moreover, we do not consider cases where a malicious Prover compromises
a Verifier's terminal or a Prover colludes with a Verifier, as the Verifier can always opt to 
not execute the protocol at all. 

\section{Design \& Implementation}\label{sec:design}
In this section, we outline our system architecture and discuss our design choices.
\let\thefootnote\relax\footnote{Demonstration videos can be found at: \url{https://tiny.cc/WAP_demos} and a sample implementation here: \url{https://github.com/alan-turing-institute/grID}.}

\begin{figure}%
    \centering
    \subfloat[]{{\includegraphics[width=0.25\textwidth]{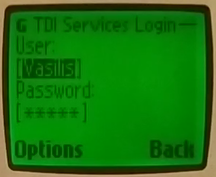} }}%
    \qquad
    \subfloat[]{{\includegraphics[width=0.25\textwidth]{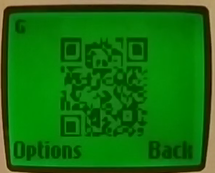} }}%
    \caption{Service login screen and QR Code as displayed on a Nokia 3510 (2002) microbrowser.}
    \label{fig:phone}%
\end{figure}

\subsection{Protocol}\label{sec:protocol}
\begin{figure}
\caption{\textbf{Authentication protocol.} The feature phone user (the Prover) authenticates to the user of a smartphone (the Verifier) 
by fetching and displaying a (single-use) QR code on the screen of their phone. The QR code token may encapsulate a digital signature, 
a URI or arbitrary other data. The recipient decodes the token and verifies the validity of the contents (if applicable) either locally 
or through a trusted online service.}
\centering
\includegraphics[width=0.90\textwidth]{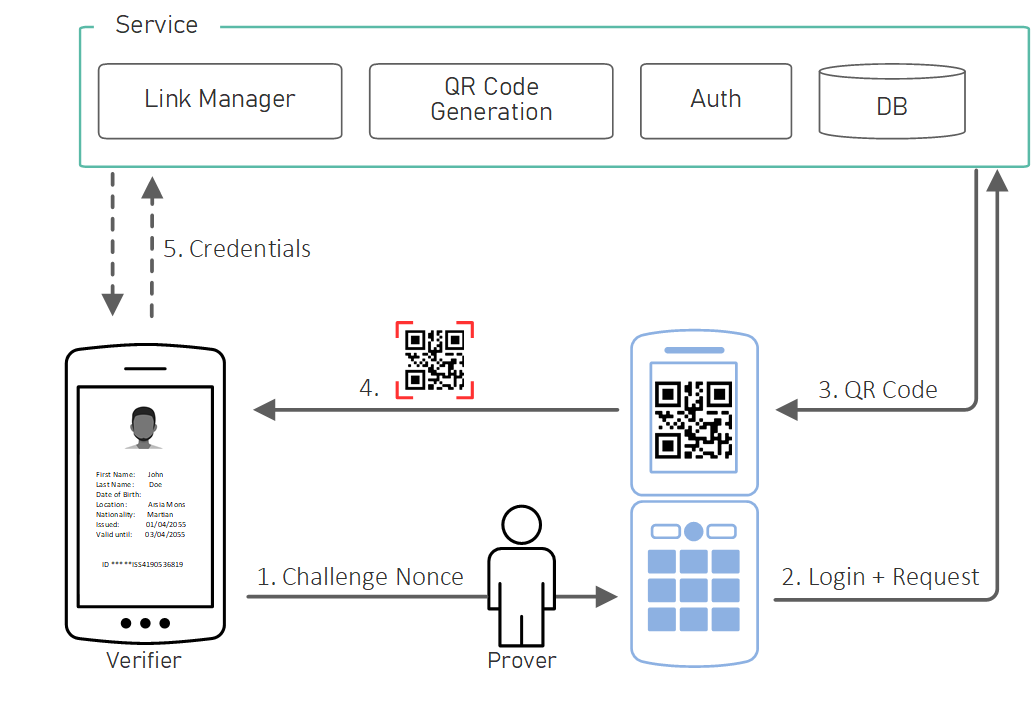}
\label{fig:protocol}%
\end{figure}

As seen in Figure~\ref{fig:protocol}, the Prover is first given a cryptographic nonce from the Verifier (step 1). This nonce serves as a challenge to prevent replay attacks and is relatively short (6-10 digits) as the Prover manually types it into the feature phone. Following the nonce input and a successful user login (Figure~\ref{fig:phone}b) to the online service, the Prover submits a request for a token (step 2). The service responds to the request and the feature phone's microbrowser displays the QR code (step 3). Subsequently, the Verifier scans and decodes the QR code (step 4). Following this step, the token verification may be carried out either online or using offline credentials.\\

\textit{Offline Credentials.}
The Verifier checks the integrity and validity of the token locally on their device (e.g., by verifying its digital signature).
In particular, they ensure that the data is signed under the online service's public key along with the correct nonce.
If the verification is successful, the Verifier accepts the Prover's claim (use case \#1 in Section~\ref{sec:usecases}).

Note that in cases such as the Covid-19 certificates \cite{Vaccine_Cert_Interim}, the tokens do not need to be ``fresh'' and thus the nonce step can
be omitted. The advantage of the nonce-free approach is that the feature phone user can use the microbrowser's cache to request tokens ahead of the time and use them later even in areas with no network coverage. In contrast, if the tokens, need to be newly-issued, the Verifier can request for the random nonce to be incorporated in the generated token.\\

\textit{Online Verification.}
Alternatively, the Verifier may use the trusted online service to fetch the user data. In this case, the token encoded in the
QR code is simply a (single-use) URI that allows the Verifier to load the user's data from the service. To prevent replay
attacks, where a user presents the same token several times the service should allow each URI to be used only once (use case \#2 in Section~\ref{sec:usecases}). 
Note that this protocol does not prevent the Verifier from launching relay attacks and impersonating the Prover to other Verifiers. 
More specifically, a malicious Verifier could use a delegate who will initiate a transaction with a victim Verifier and relay their 
nonce. In our use case, this is not a concern as the Verifier will not be able to claim government subsidy twice (the online service prevents token reuse).

\section{Related Work}
There are only a small number of digital identity and authentication applications for feature phones. Probably the most widely used example~\cite{Ma19_SMS_Study}, although one which is increasingly being replaced owing to vulnerabilities~\cite{lei2021SMSinsecurity,YooSMS15}, is the use of OTP codes as an authentication factor. These are usually sent to a mobile number which is registered to the account which a user is attempting to access. A notably more sophisticated application is Mobile-ID~\cite{mobile_id_Laud_09,imza_krimpe_14}, a solution deployed in both in Estonia and Azerbaijan which makes use of a mobile phone SIM card as a government identity document. A special-purpose Mobile-ID SIM allows mobile devices to be used for access to e-Government services and to digitally sign documents. Unlike our solution, which focuses on the in-person presentation of signed user credentials, Mobile-ID is designed for remote authentication to online services. In Mobile-ID, the SIM card is used to compute a secure digital signature on a short verification code which is shown to the user through their web browser. Laud et al.~\cite{mobile_id_Laud_09} formally model and prove the security of Mobile-ID using the Proverif protocol verification tool. 

Concerning feature phone authentication for payments, Baqer et al.~\cite{baqer_digitally_17} describe an offline payment protocol for feature phones based on exchanging short human-readable codes between payee and recipient. The authors evaluate the usability of their system with 19 participants in Nairobi who report positively upon the usability and perceived security of the system. Notably, users would visually display their authentication code to the recipient rather than to read it out loud, supporting the notion that machine-readable barcodes could be preferred. Baqer et al. also provide a more technical consideration of the short authentication codes, security and usability of their design~\cite{baqer_SMAPS_17}. Earlier work by Panjwani et al.~\cite{panjwani10} examines an Indian mobile banking service, Eko, and highlights a number of vulnerabilities as well as suggesting how to fix the existing scheme.

\section{Conclusions}
We have presented a previously overlooked approach that enables support for digital identity credentials in legacy mobile devices.
Future works could further advance this direction with the use of SIM cards as a hardware root of trust for more secure
authentication applications on all types of mobile devices. Such a development would greatly assist the efforts
(e.g., smart-feature phones with KaiOS) to bring Internet access to the next billion globally. 

\section{Acknowledgments}
This work was done for The Alan Turing institute. This work was supported, in whole or in part, by the Bill \& Melinda Gates Foundation [INV-001309]. Under the grant conditions of the Foundation, a Creative Commons Attribution 4.0 Generic License has already been assigned to the Author Accepted Manuscript version that might arise from this submission.

\bibliographystyle{splncs04}
\bibliography{bibliography}
\end{document}